\documentclass[pre,twocolumn]{revtex4}

\usepackage{graphicx,amssymb,amsmath}
\usepackage{color}
\usepackage{multirow}
\usepackage{ulem}

\newcommand{\be}{\begin{equation}}
\newcommand{\ee}{\end{equation}}

\newcommand{\bea}{\begin{eqnarray}}
\newcommand{\eea}{\end{eqnarray}}

\def\eq#1{Eq.~(\ref{#1})}
\def\eqs#1#2{Eqs.~(\ref{#1},\ref{#2})}

\usepackage{array}
\newcolumntype{C}[1]{>{\centering\let\newline\\\arraybackslash\hspace{0pt}}m{#1}}
\newcolumntype{L}[1]{>{\raggedright\let\newline\\\arraybackslash\hspace{0pt}}m{#1}}
\newcolumntype{R}[1]{>{\raggedleft\let\newline\\\arraybackslash\hspace{0pt}}m{#1}}

\begin{document}

\title{Role of charge regulation and flow slip on the ionic conductance of nanopores:\\ an analytical approach}

\author{Manoel Manghi$^\dag$}
\email{manghi@irsamc.ups-tlse.fr}
\author{John Palmeri$^\ddag$}
\author{Khadija Yazda$^\ddag$}
\author{Fran\c cois Henn$^\ddag$}
\author{Vincent Jourdain$^\ddag$}
\affiliation{$^\dag$Laboratoire de Physique Th\'eorique, IRSAMC, Universit\'e de Toulouse, CNRS, UPS, Toulouse, France}
\affiliation{$^\ddag$Laboratoire Charles Coulomb (L2C), UMR 5221 CNRS-Universit\'e de Montpellier, F-34095 Montpellier, France}

\date{\today}

\begin{abstract}
The number of precise conductance measurements in nanopores is quickly growing. In order to clarify the dominant mechanisms at play and facilitate the characterization of such systems for which there is still no clear consensus, we propose an analytical approach to the ionic conductance in nanopores that takes into account (i)~electro-osmotic effects, (ii)~flow slip at the pore surface for hydrophobic nanopores, (iii)~a component of the surface charge density that is modulated by the reservoir $p$H and salt concentration $c_s$ using a simple charge regulation model, and (iv) a fixed surface charge density that is unaffected by $p$H and  $c_s$. Limiting cases are explored for various ranges of salt concentration and our formula is used to fit conductance experiments found in the literature for carbon nanotubes. This approach permits us to catalog the different possible transport regimes and propose an explanation for the wide variety of currently known experimental behavior for the conductance \textit{versus} $c_s$.
\end{abstract}

\maketitle

\section{Introduction}

The transport of fluids in small section nano-channels and nanopores is very different from that in the bulk. Indeed, at this nanometer scale, the pore surface influences drastically the transport properties, which therefore provide a handle for characterizing the surface-fluid interactions whose range can reach throughout the whole pore section (as opposed to a thin boundary layer in the case of large diameters).
Societally important examples  motivating a growing interest in such systems arise from the search for selective and energy efficient membranes for sea water desalination and electrokinetic energy conversion (using pressure or salt concentration gradients).
The recent development of nanofluidics has led to a huge amount of experimental data on ionic transport properties in nanopores.
Among the various flux measurements, the ionic conductance $G=I/V$, where $I$ is the measured electrical current and $V$ the applied voltage, is of central interest to characterize ion transport and ion selectivity in nanopores.
Although in the last decade ionic conductance has been measured in numerous nanopores and nanochannels, including carbon nanotubes (CNT)~\cite{Strano2010,Lindsay2011,Strano2013,Noy2014,Secchi2016,Amiri2017,Yazda2017}, boron nitride nanotubes (BNNT)~\cite{Siria2013}, PDMS-glass~\cite{Strano2013b} and polymeric track-etched nanopores~\cite{Siwy2010,SciRep}, it is still not clear what role mechanisms like surface charge regulation~\cite{Ninham,RPod,Lint} and fluid slip~\cite{SciRep,Joly04,Joly06,Huang,Bieshpp16} play in determining it. The essential difference between this recent work and the early pioneering experimental and modeling studies (see, e.g., ~\cite{Koh,Westermann}) lies in the use of single well characterized nanopores, which facilitates enormously modeling.

Although the chemical nature of these various nanopores can differ a lot, some features, due to the nanoscale transport, are common and a simple theoretical model that rationalizes them is still missing.
To model these experimental results and therefore extract important nanopore characteristics such as the radius or  surface charge density, either a simple interpolation formula is used~\cite{Siria2013,Strano2013b} or the full space-charge model (Poisson-Nernst-Planck (PNP) and Stokes equations)~\cite{Morrison,Gross,Fair,Schoch2008} is solved numerically~\cite{Koh,Westermann,Lindsay2011,Biesheuvel2016,Huang,Bieshpp16}.
Recently a formula has been proposed for the conductance of nanopores bearing a constant surface charge density with or without fluid slippage at the nanopore surface~\cite{SciRep}. It has been shown that for hydrophobic nanopores, such as CNTs or polymer track-etched ones, the electro-omostic contribution can play a non-negligible role, especially for highly charged nanopore surfaces and/or strong slippage.

Some properties of the conductance at low salt concentration $c_s$ can be understood in terms of the surface charge density of the nanopore, $\sigma$, since electroneutrality imposes that the concentration of oppositely charged carriers (the counter-ions) in the pore be proportional to $\sigma$. This is the reason why a plateau is often observed for $G$ at low $c_s$ for a constant surface charge density.
However, Pang \textit{et al.}~\cite{Lindsay2011} measured a non-constant conductance $G\propto c_s^{\alpha}$ with $\alpha\simeq0.37$ at low $c_s$ in CNTs. Later Secchi \textit{et al.}~\cite{Secchi2016} showed that the CNT conductance at low $c_s$ varies with $c_s$ and the $p$H of the solution and extracted an exponent $\alpha=1/3$. A similar observation has been made recently by Yazda \textit{et al.}~\cite{Yazda2017}, but with a different power law.
The theoretical explanation proposed in the literature is \textit{charge regulation}~\cite{Secchi2016,Biesheuvel2016}, i.e. charged groups appear at the nanopore surface when the $p$H is increased, which could be due various mechanisms, such as the adsorption of hydroxyde ions~\cite{Secchi2016} or the dissociation of carboxylic groups COOH. This adsorption/dissociation is also influenced by screened electrostatic interactions, the screening being due to the presence of ions in the pore. This could be the reason why the behavior of $G$ vs. $c_s$ is modified at low $c_s$. Since Secchi \textit{et al.}~\cite{Secchi2016} and Biesheuvel and Bazant~\cite{Biesheuvel2016} each developed different approximate theoretical approaches that led to different exponents ($\alpha=1/3$ and 1/2, respectively) at low salt concentration, the situation needs to be reexamined. Furthermore, in addition to charge regulation, the role of flow slip, neglected in~\cite{Secchi2016,Biesheuvel2016}, but already touched upon in~\cite{Bieshpp16}, needs to be assessed. In this last reference the influence of slip on conductivity was investigated, but only for the single case of a relatively large nanopore radius (5 nm) and small slip length (1.25 nm). The conclusion in this special case was that although slip leads to an increase in conductivity, the effect is minor. It is known from theory and simulations, however, that slip lengths in hydrophobic nanopores can be very large~\cite{Vino}, and, as observed in molecular dynamics simulations, can even be much larger than the pore radius~\cite{Barrat,Secchi2016b}. Such large slip lengths can modify in important ways the transport properties of hydrophobic nanopores~\cite{SciRep,Joly04,Joly06,Huang,Bieshpp16}. These effects for slippery hydrophobic surface can be modified if the mobility of surface charges is taken into account, as shown for electro-osmotic flow in~\cite{Maduar}.

Nevertheless it should be stressed that the nature of the surface charge of CNTs has not yet been elucidated. It can have \textit{a priori} several origins, which can be intrinsic to the nanopore, such as an affinity of the graphene surface for OH$^-$ ions or structural defects creating local charges (crystallographic defects or weak acid groups, e.g. COOH), but also extrinsic, such as  chemical or electrostatic doping induced by the nanotube environment (resin, matrix, etc.).

Interesting recent work has attempted to go beyond the standard Space Charge model by using input from molecular dynamics simulations to introduce inhomogeneous dielectric function and viscosity profiles. The experimentally observed excess surface conductivity can be explained for both hydrophilic and hydrophobic surfaces using the non-electrostatic ion-surface interaction as a fitting parameter~\cite{Bonthuis12,Bonthuis13}. Other recent work has incorporated a more detailed description of  aqueous interfaces into the Space Charge model by employing  a basic Stern model to describe the uncharged surface dielectric layer (in conjunction with slip in ~\cite{Bieshpp16}) and more sophisticated extensions for specific surfaces, such as an electrical triple-layer model for silica to include the surface contributions from salt ions~\cite{Wang}.

In this paper, we propose an extension of the analytical formula for nanopore conductance (based on the Space Charge model incorporating slip) that we  proposed in~\cite{SciRep} to include the charge regulation mechanism. In an effort to increase the physical content of the model incrementally and retain as much simplicity as possible, we do not attempt to include the additional features evoked above (such as surface ion mobility, Stern layer effects, etc., cf.~\cite{Maduar,Bonthuis12,Bonthuis13,Wang,Bieshpp16}). Although the approach presented here can be extended to include these additional features, we believe that the next important step in understanding transport in CNTs is to combine charge regulation and flow slip. Our analytical formula is then used to fit experimental conductance measurements in CNTs found in the literature. We concentrate here on small radius CNTs with large slip lengths for which the ratio of slip length to pore radius is very large (as expected for such nanopores~\cite{Secchi2016b}). We show in particular that (i) slip can play an important and even dominant role and (ii) to get  physically reasonable fits to the conductivity data for the tightest nanopore studied in~\cite{Secchi2016} (3.5 nm radius) over the whole $p$H range investigated experimentally, a weak fixed surface charge density must be included along with slip and surface charge regulation. Previous efforts to fit the data for this CNT simultaneously for the four studied values of $p$H using just surface charge regulation were not successful, even with unphysically high values of maximum surface charge density (a difficulty that led the authors of~\cite{Biesheuvel2016} to suggest that slip may play an important role). Our analytical approach is complementary to the one based on numerical solutions to the PNP equations and can be used to draw a global ``phase diagram"  for the conductance mechanisms in the salt concentration vs. surface charge density plane, which is less accessible by purely numerical methods (cf.~\cite{Huang,Biesheuvel2016,Bieshpp16}).

\section{Model}

We consider a monovalent salt (such as NaCl or KCl) in an aqueous solution inside a cylindrical nanopore of radius $R$ and length $L$. We assume that $L\gg R$ so that the ionic concentration in the pore is independent of the distance $z$ along the cylinder axis and end effects are negligible. At both ends the nanopore is connected to two electrolyte reservoirs at salt concentration $c_s$.

\begin{figure}[t]
\begin{center}
\includegraphics[width=9cm]{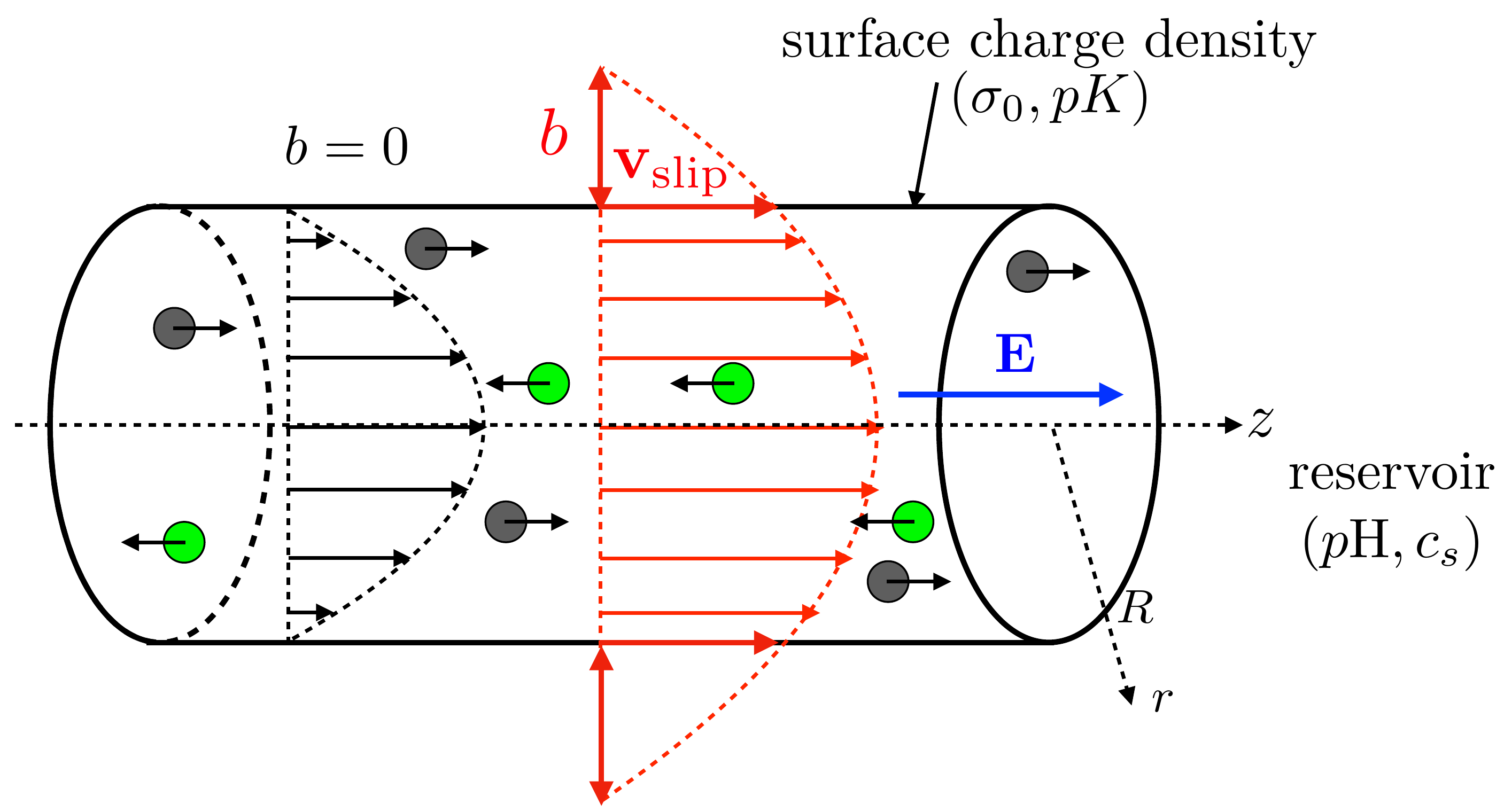}
\end{center}
\caption{Sketch of the nanopore showing counterions (black spheres) and co-ions (green/light spheres) for a negative surface charge density $\sigma$ which is controlled by its saturation value $\sigma_0$, the $pK$ of the anion adsorption or neutral group dissociation at the surface, and the $p{\rm H}$ and the salt concentration $c_s$ in the reservoirs. A constant electric field is applied $\mathbf{E}$ is applied between the two reservoirs located at the pore extremities. Two velocity fields are sketched, with (in red) or without (in black) slippage at the pore surface (the slip length is noted $b$).}
\label{sketch}
\end{figure}

We assume that a negative surface charge develops following a simple charge regulation mechanism~\cite{Ninham,RPod,Lint} whose detailed physical chemistry we leave unspecified awaiting further studies on this topic. This model is general enough to include (i)~anion (e.g. OH$^-$) adsorption at a hydrophobic nanopore surface, and (ii)~surface grafted acid group (e.g. COOH) dissociation, leading to  the release of hydronium ions H$_3$O$^+$ in the reservoir. Note that the model is easy to modify in the case of the formation of positive surface groups (such as NH$_3^+$). As a first hint into this complex problem, we follow the usual practice of neglecting dielectric effects~\cite{PRL,JCP1} and ion-ion correlations~\cite{JCP}, which could potentially play an important role, and thus treat the electrostatic statistical problem at the level of the mean-field Poisson-Boltzmann (PB) equation.

Our goal is to obtain the variations of the nanopore conductance as a function of the bulk salt concentration $c_s$ by assuming that the conductivity (at low $c_s$) is influenced by a nanopore surface charging mechanism, for example through anion adsorption or neutral group dissociation.

\begin{figure}[t]
\begin{center}
\includegraphics[width=6cm]{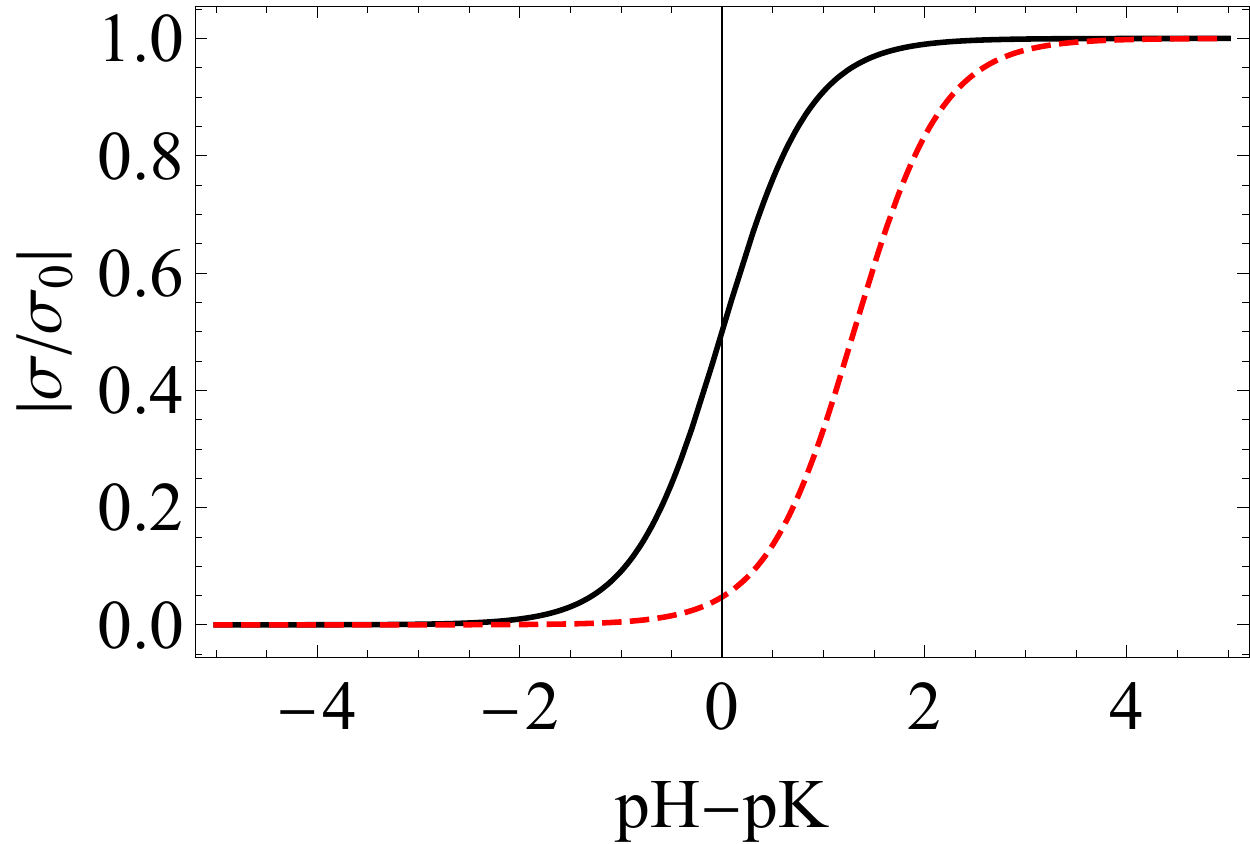}
\end{center}
\caption{Absolute value of the dimensionless surface charge density versus $p{\rm H}-pK$ for $|\phi_s|=0$ (black) and 3 (red dashed) [\eq{CRM2}]. Decreasing the salt concentration increases the surface potential at constant $p{\rm H}$ and favors the formation of neutral groups. For sufficiently large $p$H $\gg pK$, $\sigma$ remains close to saturation ($\sigma_0$) until a low threshold concentration is reached.}
\label{sig}
\end{figure}
Following the usual charge regulation (or Langmuir) model in its simplest form, the (negative) surface charge density is taken to have an absolute value
\be
\sigma= \frac{\sigma_0}{1+10^{pK-p{\rm H}}\  e^{|\phi_s|}}
\label{CRM2}
\ee
where $pK$ refers to the equilibrium constant of the charging mechanism, $p{\rm H}\equiv -\log[{\rm H_3O}^+]_b$ refers to the external (bulk) reservoirs,  $\sigma_0= n e/(2\pi RL)$ with $n$ ionizable groups ($e$ is the positive elementary  charge), and $\phi_s=e\psi_s/k_BT  \le 0$ is the dimensionless electrostatic potential at the pore surface.
An important limitation to \eq{CRM2} should be kept in mind: the acid, such as HCl, or base, such as NaOH, added to the bulk to adjust the $p{\rm H}$ must be low enough in concentration to avoid modifying the surface potential $\phi_s$, which in the approximation adopted here is fixed entirely by  bulk salt (with the ions coming from the added acid or base playing the role of a spectators, or trace, species). Within this approximation, which is valid when
\be
-\log c_s \ll p{\rm H} \ll 14 + \log c_s
\label{cr}
\ee
(with $c_s$ in mol/L) the value of the surface charge density depends only on the effective (salt concentration dependent) $p{\rm H}$ value in the pore: $p{\rm H}_{\rm pore} = p{\rm H}-|\phi_s (c_s)|/\ln(10)  \leq   p{\rm H}$, which decreases with increasing $c_s$.
In Fig.~\ref{sig} we plot $\sigma/\sigma_0$ as a function of $p{\rm H}-pK$ and recall that  $\sigma/\sigma_0$ goes quickly to its maximum value for $p{\rm H}_{\rm pore}-pK > 1$ and to 0 for $p{\rm H}_{\rm pore}-pK < 1$. At high reservoir salt concentration, the pore charge is screened, $|\phi_s|$ goes to zero, and therefore $\sigma$ reaches its maximum $p{\rm H}$ dependent value,
\be
\sigma_{\rm max}(p{\rm H}) = \frac{\sigma_0}{1+10^{pK-p{\rm H}}} \le \sigma_0.
\label{CRMb}
\ee
As the reservoir salt concentration is lowered, more and more co-ions are excluded and $|\phi_s|$ increases in order to maintain electro-neutrality in the pore, driving the pore to lower and lower charge states (see, e.g.,~\cite{Secchi2016,Biesheuvel2016}):
\be
\sigma \approx \sigma_0 \ 10^{p{\rm H}-pK} e^{-|\phi_s|}.
\label{crsc}
\ee

To relate the dimensionless electrostatic potential $\phi_s=\phi(r=R)$ (where $r$ is the radial coordinate in the pore) to the surface charge density and therefore obtain an implicit relation which gives $\sigma$ as a function of the salt concentration $c_s$ and the $p$H of the bulk solution, we would need to solve the PB equation in a cylindrical pore with the appropriate boundary conditions:
\bea
\frac1{r}\frac{\partial}{\partial r}\left(r\frac{\partial \phi}{\partial r}\right)=\frac1{\lambda^2_{\rm DH}}\sinh\phi\label{PBE}\\
\left.\frac{\partial \phi}{\partial r}\right|_{r=0}=0,\quad \left.\frac{\partial \phi}{\partial r}\right|_{r=R}=-4\pi\ell_B\frac{\sigma}e,
\eea
where $\lambda_{\rm DH} =(8\pi\ell_B c_s)^{-1/2}$ is the Debye screening length in the bulk, and $\ell_B=e^2/(4\pi\epsilon_0\epsilon k_BT)$ the Bjerrum length, equal to 0.7~nm  in water at room temperature (recall that $\sigma$ is the absolute value of the negative surface charge).

\begin{figure}[t]
\begin{center}
\includegraphics[width=8cm]{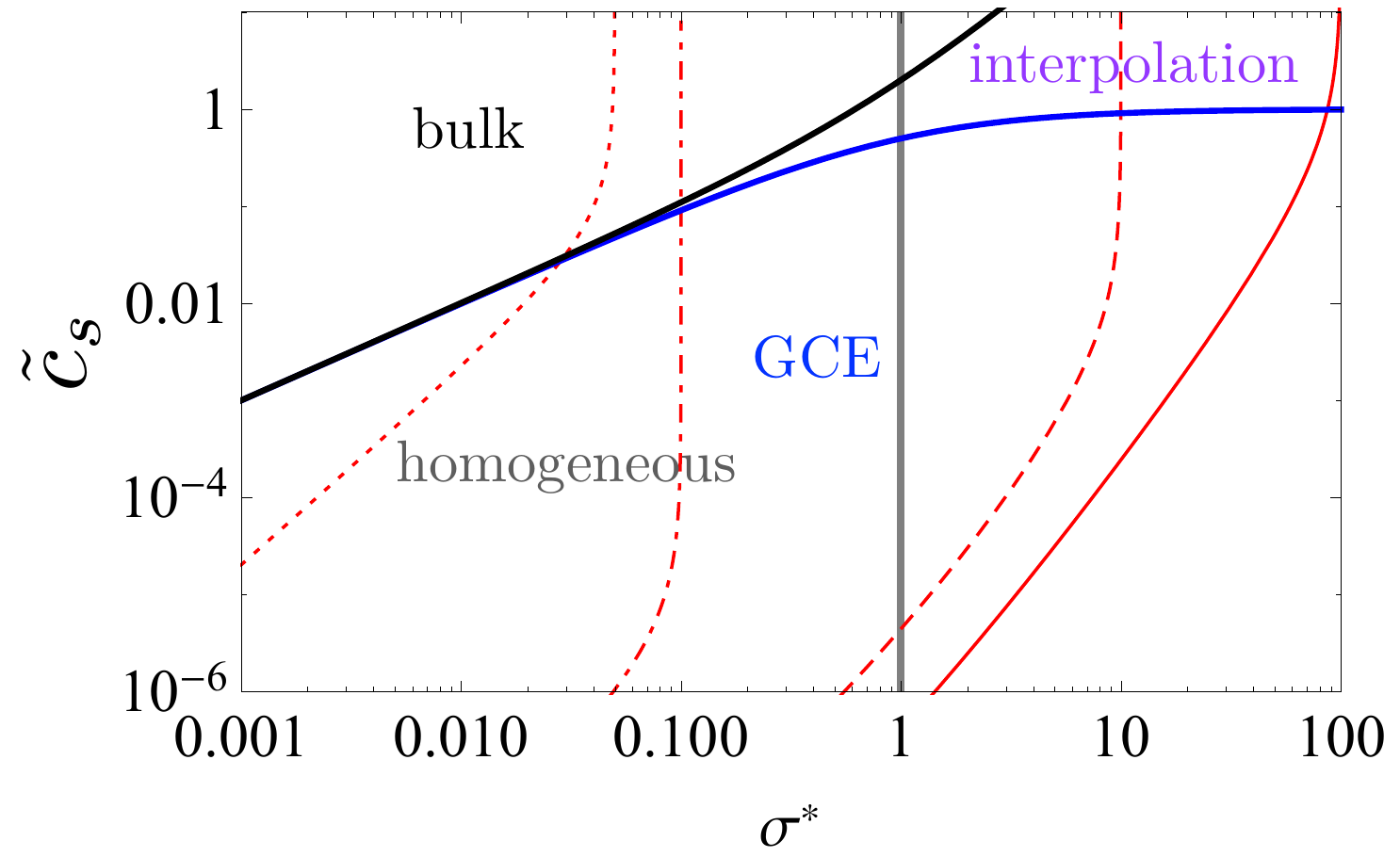}
\end{center}
\caption{Diagram in the $(\sigma^*,\tilde{c}_s)$ plane of the various regimes (see text): {\it bulk} above the black line [\eq{c_bulk}],  {\it homogeneous} for $\sigma^*<1$, {\it GCE} below the blue line [\eq{GCE}], and the remaining {\it interpolation} one. The four thin red curves correspond to $\sigma^*(c_s)$ for (from left to right) $(\sigma^*_0,p{\rm H}-pK)=(0.1,0)$(dotted), (0.1,5) (dashed-dotted), (10,5) (dashed), and (100,5) (solid).}
\label{diag}
\end{figure}

Although no exact solution of \eq{PBE} is known for arbitrary pore radius, surface charge density, and bulk concentration, certain limiting cases lead to relatively simple approximations, as summarized in \cite{SciRep} (see Fig.~\ref{diag}).\\
(i)~In the homogeneous approximation, the electrostatic potential is taken as constant over the pore cross-section. This approximation,
\be
\label{HE}
e^{|\phi_s|} \approx e^{|\phi_{\rm H}|} = \frac{|\sigma|}{e R c_s}\left[\sqrt{1 + \left(\frac{e R c_s}{\sigma}\right)^2} + 1 \right],
\ee
is valid over the whole concentration range for $\sigma^*<1$, where the dimensionless surface charge density is
\be
\sigma^* = \sigma \frac{\pi R \ell_B}e.
\ee
(ii) In the good co-ion exclusion (GCE) limit,
\be
e^{|\phi_s|} \approx e^{|\phi_{\rm GCE}(R)|}=16
\sigma^*(1 + \sigma^*)
\left(\frac{\lambda_{\rm DH}}{R}\right)^2,
\label{phiGCE}
\ee
This approximation is valid if the normalized GCE electrostatic potential at the pore center is greater than 1 (see for example Eq.~(10) of \cite{SciRep}):
\be
|\phi_{\rm GCE} (0)|
=\ln \left[16
\frac{\sigma^*}{1 + \sigma^*}
\left(\frac{\lambda_{\rm DH}}{R}\right)^2 \right] >1,
\ee
that is for
\be
\tilde c_s < \tilde c_{\rm GCE}\equiv\frac{\sigma^*}{1+\sigma^*},
\label{GCE}
\ee
where
\be
\tilde c_s = \pi \ell_B R^2 c_s =\frac{1}{8}\left(\frac{R}{\lambda_{\rm DH}}\right)^2
\ee
is a dimensionless salt concentration. At low surface charge density $\sigma^* < 1$,  the homogeneous approximation is valid and $|\phi_{\rm GCE} (0)| \approx|\phi_{\rm GCE} (R)|$. At very high surface charge density $\sigma^* \gg 1$, and although $|\phi_{\rm GCE} (0)|$  saturates to $\ln (2/\tilde c_s)$ (for $\tilde c_s <\tilde c_{\rm GCE} \approx 1$), $|\phi_{\rm GCE} (R)|$ grows indefinitely.

We can combine these two approximations by using the following interpolation formula:
\be
e^{|\phi_s|} \approx \frac{\sigma^*(1 + \sigma^*)}{\tilde c_s}\left[\sqrt{1 + \left[\frac{\tilde c_s}{\sigma^*(1 + \sigma^*)}\right]^2} + 1 \right].
\label{interpol}
\ee
Equation~(\ref{interpol}) allows us to obtain the following correct three limits: (i) the high concentration (bulk) one,
\be
\tilde c_s \gg \tilde c_{\rm bulk}\equiv\sigma^*(1+\sigma^*)
\label{c_bulk}
\ee
where $\phi_s\to 0$; (ii) the weak surface charge homogeneous one, $\sigma^* < 1$; and (iii) the low concentration GCE one, $\tilde c_s < \tilde c_{\rm GCE}$ [when $\sigma^* < 1$ we enter the homogeneous GCE regime, which overlaps with (ii)]. Equation~(\ref{interpol}) should therefore be a good approximation for $\phi_s$ over the whole range of pore surface charge density and bulk salt concentration.

The different regimes in the $(\sigma^*,\tilde{c}_s)$ plane presented above are shown in Fig.~\ref{diag}. The interpolation regime that exists at sufficiently high $\sigma^*$ for intermediate $\tilde{c}_s$ is characterized by a high dimensionless electrostatic potential near the pore surface (with corresponding good co-ion exclusion) and a low potential near the pore center (with corresponding poor co-ion exclusion). This regime is therefore an intermediate one in terms of ion selectivity.
The low surface charge limit of charge regulation, \eq{crsc}, is valid in the GCE regime [see \eqs{phiGCE}{GCE}] whenever
\be
\tilde c_s<\tilde c_{\rm CR}\equiv2\sigma^*(1+\sigma^*)10^{pK-p{\rm H}}.
\label{scaling}
\ee

Following the theoretical framework developed in~\cite{SciRep}, the nanopore conductance is
\be
G=\frac{\pi R^2}{L} \kappa
\ee
where the conductivity is given by
\bea
\kappa&=&e^2c_s \left[ \mu_+\langle e^{-\phi}\rangle +\mu_-\langle e^{\phi}\rangle \right] \nonumber\\
&&+\frac{e^2 c_s}{2\pi\ell_{B}\eta}\langle (\phi(R)-\phi)\sinh\phi\rangle+\kappa_{\rm slip}
\label{kappa1}
\eea
with $\eta=8.94 \times10^{-4}$~Pa.s the water viscosity and $\mu_i$  the  mobility of ion $i$ (in the absence of better values in CNTs, it is taken equal to its bulk value) and $\langle A\rangle=\frac2{R^2}\int_0^R A(r) r dr$ corresponds to average quantities in the pore. The first  term on the rhs of \eq{kappa1} gives the ionic migration contribution, whereas the second and third terms give the electro-osmotic one.
The last term,
\be
\kappa_{\rm slip}=\frac{2\sigma^2b}{\eta R},
\label{slip}
\ee
is the exact slip contribution to the electro-osmotic one (see Appendix) and comes directly from the non-vanishing solvent velocity at the pore wall, ${\bf v}_{\rm slip}=-\frac{b\sigma}{\eta}{\bf E}$ where ${\bf E}$ is the applied electric field and $b$ the slip length (taken to be constant, i.e.  independent of $\sigma$ and $R$). Following~\cite{SciRep}, using an interpolation similar to the one that led to \eq{interpol}, \eq{kappa1} simplifies to
\bea
\kappa &=& e^2 (\mu_+ +\mu_-)c_s \sqrt{1+\left(\frac{\sigma}{eRc_s}\right)^2} + \frac{e|\sigma|}{R}(\mu_+ -\mu_-)\nonumber \\ &+& \frac{\sigma^2}{2\eta}\left[\frac2{\sigma^*}\left(1-\frac{\ln(1+\sigma^*)}{\sigma^*}\right)+4\frac{b}{R}\right].
\label{kappa}
\eea
The first two terms on the rhs of \eq{kappa} give the ionic migration contribution, whereas the third term gives the electro-osmotic one, including the exact slip contribution.

We implicitly suppose that the proton concentration in the pore, coming from the acid introduced into the bulk to adjust the $p{\rm H}$ to low values, does not substantially affect the electrical migration contribution despite the very high proton mobility (5 times higher than that of K$^+$  and 7 times higher than that of Na$^+$ in the bulk). The increase in counter-ion concentration coming from adding a base such as NaOH to adjust the $p{\rm H}$ to high values is also assumed negligible [see \eq{cr}].

Equation~(\ref{kappa}) has been successfully used to fit conductivity experiments  in hydrophobic track-etched  nanopores without charge regulation~\cite{SciRep}.

To simplify the analysis we use the following additional dimensionless quantities:
\be
\tilde \kappa = \kappa \frac{2\pi^2R^2\ell_B^2 \eta}{e^2}; \quad \tilde\mu_i = \mu_i 2\pi\eta \ell_B ;\quad \tilde b = \frac{b}{R}
\ee
For a typical nanopore of radius $R=1$~nm, we have $\tilde c_s=1.2\,c_s [{\rm mol/L}]$. Note, moreover, that with this choice, $\tilde \kappa$ is also the dimensionless conductance defined as $\tilde G=(2\pi\ell_B^2L\eta/e^2)G=\tilde \kappa$.
\eq{CRM2}, \eq{interpol} and \eq{kappa} therefore simplify to
\bea
\tilde\kappa &=& (\tilde\mu_+ +\tilde\mu_-)\tilde c_s \sqrt{1+\left(\frac{\sigma^*}{\tilde c_s}\right)^2} + \sigma^*(\tilde \mu_+ -\tilde \mu_-) \nonumber \\ &+& 2\sigma^*\left(1-\frac{\ln(1+\sigma^*)}{\sigma^*}\right)+4\tilde b \sigma^{*2}\label{KCRM}\\
\tilde c_s &=&
\frac{ 2h\sigma^{*2} (1+\sigma^*)(\sigma_0^*-\sigma^*) }{ [\sigma_0^*- (1+h) \sigma^*] [\sigma_0^*- (1-h) \sigma^*] }
\label{CR}
\eea
where $h=10^{pK-p{\rm H}}$ (and $\sigma_0^*=\sigma_0\pi R\ell_B/e$). This system of coupled non-linear equations, which can be simply solved parametrically as a function of $\sigma^*$ (without any need for numerical methods), is the central result of our paper. In the following we first discuss the various regimes of $\tilde \kappa(\sigma^*)$ as a function of $\tilde c_s(\sigma^*)$ ($0<\sigma^*<\sigma^*_{\rm max}$) and then use this formula to fit some experimental data found in the literature.
Equation~(\ref{CR}) can be used to trace $\tilde c_s$ \textit{vs.} $\sigma^*$ in the $(\sigma^*,\tilde{c}_s)$ plane (Fig.~\ref{diag}). There are four distinct cases:
(i)~pure scaling, $\sigma^*\propto \tilde{c}_s^\beta$, in the low surface charge GCE regime ($\sigma^*<1$) for low saturation surface charge density $\sigma_0$ and low $p$H (dotted red curve in Fig.~\ref{diag});
(ii)~a constant surface charge density plateau in the low surface charge GCE regime, followed by a cross-over to scaling at low concentration for low $\sigma_0$ and high $p$H (dashed-dotted red curve);
(iii)~a constant surface charge density plateau in the high surface charge GCE regime ($\sigma^*>1$), followed by a cross-over to scaling at low concentration, first in the high surface charge GCE regime and then in the low ($\sigma^*<1$), for high $\sigma_0$ and high $p$H (dashed red curve);
(iv)~scaling, first in the high surface charge GCE regime and then in the low, for very high $\sigma_0$ and high $p$H (solid red curve).
\\

The system of equations given in \eqs{KCRM}{CR} simplifies in several scaling regimes for the conductivity, which are attained when the charge regulation relation, \eq{CRM2}, and surface potential, \eq{interpol}, tend to high or low concentration limits:
\begin{itemize}
	\item In the \textit{high salt concentration} limit where $\tilde c_s\gg \tilde c_{\rm bulk}$ (which corresponds to $|\phi_s|\to 0$), one deduces from \eq{CR} that $\sigma^*\simeq\sigma_{\max}^*(p{\rm H})=\sigma_0^*/(1+h)$ (which is the maximum absolute value of the surface charge density at a given $p{\rm H}$, see \eq{CRMb}), and we recover the bulk conductivity
\be
\tilde\kappa^{\rm bulk} \simeq (\tilde\mu_+ +\tilde\mu_-)\tilde c_s
\label{HSL}
\ee
since the surface charge effects are screened. This limit also corresponds to low $p{\rm H}\ll pK$ for which very few surface groups are ionized and $\sigma\to0$.
	\item In the \textit{low surface charge density} (homogeneous) GCE limit, reached at low salt concentration and low $\sigma^*$, $\tilde c_s < \sigma^* < 1$, we find from \eq{CR} that if the inequality~(\ref{scaling}) is satisfied, then $\sigma^{*2}\simeq \sigma_0^* \tilde c_s/(2 h)$, i.e. $\sigma\propto\sqrt{c_s}$ and the conductivity becomes
\be
\tilde\kappa^{\rm GCE} \simeq 2 \tilde\mu_+ \left( \frac{\sigma_0^* \tilde c_s}{2h} \right)^{1/2}
+ (1 + 4 \tilde b)\frac{\sigma_0^* \tilde c_s}{2h}
\label{LSL}
\ee
The leftmost two curves in Fig.~\ref{diag} at low $c_s$ and $\sigma_0$ illustrate the regime where we expect this type of behavior. The first (dominant) term at sufficiently low $\tilde c_s$ is due to electrical migration and the second (asymptotically subdominant) one is due to electro-osmosis. We therefore expect at low enough $c_s$ a scaling law with an exponent of 1/2 (in agreement with~\cite{Biesheuvel2016}) . At low but intermediate $c_s$ and high enough slip length the second term may dominate and lead to a cross-over exponent of 1.
	\item In the \textit{high surface charge density} (inhomogeneous) GCE limit, reached at low salt concentration and and high $\sigma^*$, $\tilde c_s < 1 < \sigma^*$, we find from \eq{CR} that if the inequality~(\ref{scaling}) is satisfied, then  $\sigma^{*3}\simeq \sigma_0^* \tilde c_s/(2 h)$ and therefore
\be
\tilde \kappa^{\rm inter}  \simeq  2(1+\tilde \mu_+) \left(\frac{\sigma_0^* \tilde c_s}{2h}\right)^{1/3}+ 4 \tilde b \left(\frac{\sigma_0^* \tilde c_s}{2 h}\right)^{2/3}
\label{inter}
\ee
The rightmost two curves in Fig.~\ref{diag} at intermediate $c_s$ and $\sigma_0$ illustrate the regime where we expect this type of behavior.
We thus formally expect at low enough intermediate $c_s$ a cross-over scaling law with an exponent of 1/3 (the low concentration scaling regime predicted in~\cite{Secchi2016}). Our analysis, which differs from the thin double argument proposed in~\cite{Biesheuvel2016} to explain the origin of this scaling regime, shows that it can only be an intermediate one observable at sufficiently high values of maximum surface charge density $\sigma_{\rm max}(p{\rm H})$ (for \textit{intermediate} values of low $c_s$) because charge regulation will eventually drive the system into the low surface charge density (homogeneous) GCE regime with an exponent of 1/2. The first (dominant) term at sufficiently low intermediate $\tilde c_s$ is due to both electrical migration and the non-slip electro-osmotic contribution and  the second (subdominant) one is due to the slip part of the electro-osmotic contribution.
Without slip we therefore expect an intermediate scaling behavior with an exponent of 1/3, although unphysically high values of maximum surface charge densities may be needed to actually observe this regime ($\sigma_0^*\gg1$, i.e. $\sigma\gg 0.1$~C/m$^2$ for $R\simeq 1$~nm). With sufficiently  high slip length we expect the first (putatively dominant) term to be completely masked by the second one because a sufficiently high value of the prefactor (proportional to $\tilde b$) can counterbalance the higher exponent for sufficiently low but still intermediate $\tilde c_s$. Then the intermediate scaling behavior would have an exponent of 2/3.
\end{itemize}

\begin{figure}[t]
\begin{center}
\includegraphics[width=8cm]{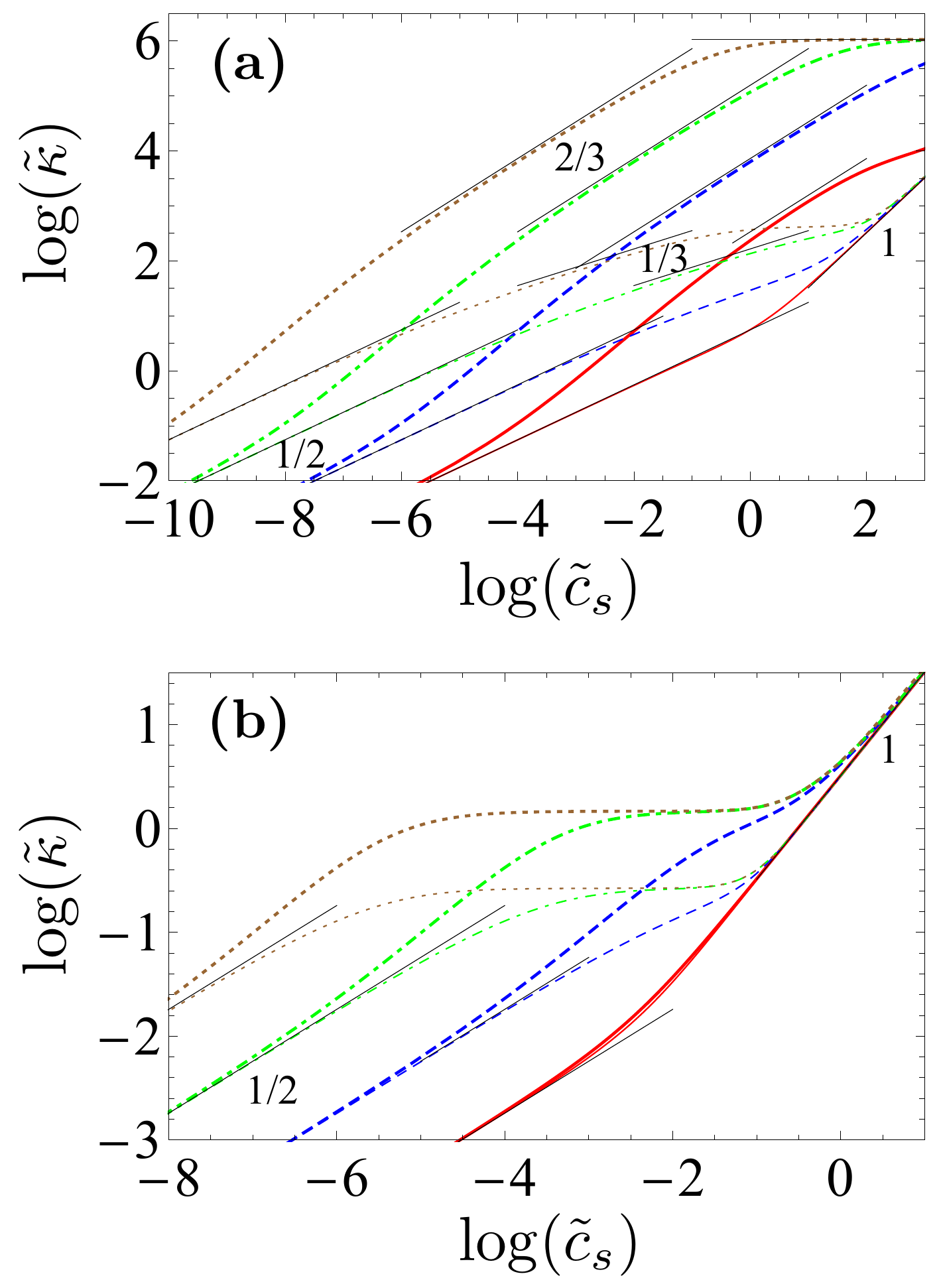}
\end{center}
\caption{Dimensionless conductivity $\tilde \kappa$ versus $\tilde c_s$ given by \eqs{KCRM}{CR} for $p{\rm H}-pK=-1,1,3,5$ (solid red, dashed blue, dotted-dashed green, dotted brown) and $\tilde b=0$ (thin curves) or $\tilde b=30$ (thick curves) (Log-Log scale): (a)~$\sigma_0^*=94$, and (b)~$\sigma_0^*=0.1$. The thin solid lines correspond to the power laws $\tilde c_s^{1/2}$ and $\tilde c_s$ of \eqs{HSL}{LSL} found in the high and low salt limit, respectively. The power laws in $\tilde c_s^{2/3}$ (for $\tilde b=30$) and $\tilde c_s^{1/3}$ (for $\tilde b=0$) appearing in (a) at intermediate $\tilde c_s$ are explained in the text.}
\label{th_plots}
\end{figure}

Before fitting the available data on conductivity in nanopores, we study theoretically the different regimes.
We thus consider a nanopore with $R=1$~nm and a NaCl electrolyte with the following mobilities $\mu_+= 3.3\times 10^{11}$~s/kg (Na$^+$) and $\mu_-=5.0\times 10^{11}$~s/kg (Cl$^-$), at room temperature ($\ell_B=0.7$~nm), which leads to the dimensionless mobilities $\tilde \mu_+=1.28$ and $\tilde \mu_-=1.95$.
\textit{A priori} three unknown parameters remain: the maximum surface charge density $\sigma_0$ attainable at high $p$H (or the number of ionizable groups $n$), the $p{\rm H}-pK$ value, and the slip length $b$ which is close to 0 for hydrophilic nanopores and can be as high as 300~nm for very hydrophobic ones (such as CNTs~\cite{Secchi2016b}).
In Fig.~\ref{th_plots} we present results for an unphysically large surface charge density of $\sigma_0=6.8$~C/m$^2$ ($\sigma_0^*=94$) in order to illustrate theoretically the various intermediate regimes discussed above, which are not all visible for physically reasonable surface charge densities (for comparison, the extremely highly charged BNNT studied in Ref.~\cite{Siria2013} showed maximum surface charge densities less than 2~C/m$^2$). Furthermore, we choose $b=0$ or 30~nm ($\tilde b=30$) as in~\cite{SciRep}.

In Figure~\ref{th_plots}(a) is shown the conductivity for various values of $p{\rm H}-pK=-1,1,3,5$. We clearly observe the various regimes presented above, the asymptotic GCE regime in $c_s^{1/2}$ at very low $c_s$ and the bulk one in $c_s$ at high $c_s$. In particular, the intermediate regime in $c_s^{1/3}$ appears only at high $p$H and high surface charge density for $b=0$ (thin dotted-dashed green and brown dotted curves at $p{\rm H}-pK=3$ and 5) in order to satisfy the inequality~(\ref{scaling}). If slippage is taken into account, a large increase of the conductivity at intermediate concentrations occurs which varies as $c_s^{2/3}$, whatever the $p$H value (thick curves).
These low and intermediate concentration scaling regimes appear for $\tilde c_s<\min(\tilde c_{\rm GCE},\tilde c_{\rm CR})$.
At large $p$H the surface charge starts to saturate to $\sigma_{\max}$ at intermediate or high $c_s$ and the conductivity plateaus to $\tilde \kappa\simeq4 \tilde b\sigma_{\max}^{*2}$ for large enough $b$, as long as the system remains in the high surface charge GCE regime (i.e., $\tilde c_{\rm CR}<\tilde c_s<\tilde c_{\rm GCE}$).

For small diameter nanopores and lower and more realistic values of $\sigma_0$, as shown in Fig.~\ref{th_plots}(b) for $\sigma_0^*=0.1$, the behavior is quite different. At low $p{\rm H}$, the surface charge $\sigma^*$ is so small over the entire salt concentration range that the conductivity interpolates directly between the two limiting behaviors, bulk and homogeneous GCE, given in \eqs{HSL}{LSL}, which vary respectively as $c_s$ and $c_s^{1/2}$ (red curves).
For increasing $p{\rm H}$ the saturation occurs at low salt concentrations and the conductivity profile becomes close to the one at constant $\sigma$ except for very low $c_s$. Indeed, when the inequality~(\ref{scaling}) is reversed ($\tilde c_{\rm CR}<\tilde c_s<\tilde c_{\rm GCE}$), the system enters the low concentration GCE (plateau) regime \textit{before} the surface charge density begins to deviate significantly from its maximum value $\sigma_{\max} \approx \sigma_0$. For low enough $\tilde c_s<\min(\tilde c_{\rm GCE},\tilde c_{\rm CR})$, however, the conductivity eventually enters the scaling regime. For such low values of $\sigma_0$ the system enters the \textit{homogeneous} GCE regime at low $c_s$ and therefore large enough flow slippage leads to a shift of the plateau value from the no-slip value $\tilde \kappa\simeq 2 \tilde \mu_+ \sigma^*_{\max}$ to the slip one, $\tilde \kappa\simeq (1 + 4 \tilde b)\sigma_{\max}^{*2}$.

Moreover, a shoulder in $\tilde \kappa(\tilde c_s)$ is observed at intermediate values of $\tilde c_s$ which leads to apparent power laws $\tilde \kappa\propto \tilde c_s^{\alpha}$ with $1/2\leq \alpha\leq 1$ (before saturation), where the value 1 for $\alpha$ corresponds to the second term in \eq{LSL}. Such high values of $\alpha$ cannot be observed without slippage at intermediate values of $\tilde c_s$, where $\alpha\leq1/2$.
A glance at Fig.~\ref{th_plots} shows that the conductivity approaches the asymptotic low concentration scaling, $\tilde c_s^{1/2}$, from above in the presence of slip and below without [except at very low saturation surface charge density/low $p$H, where the cross-over is directly from bulk to low concentration scaling (rightmost thin red curve in Fig.~\ref{th_plots}(b))]. This type of qualitative behavior can therefore be used as a signature of the presence or absence of strong slip effects.
\begin{figure}[t]
\begin{center}
\includegraphics[width=7cm]{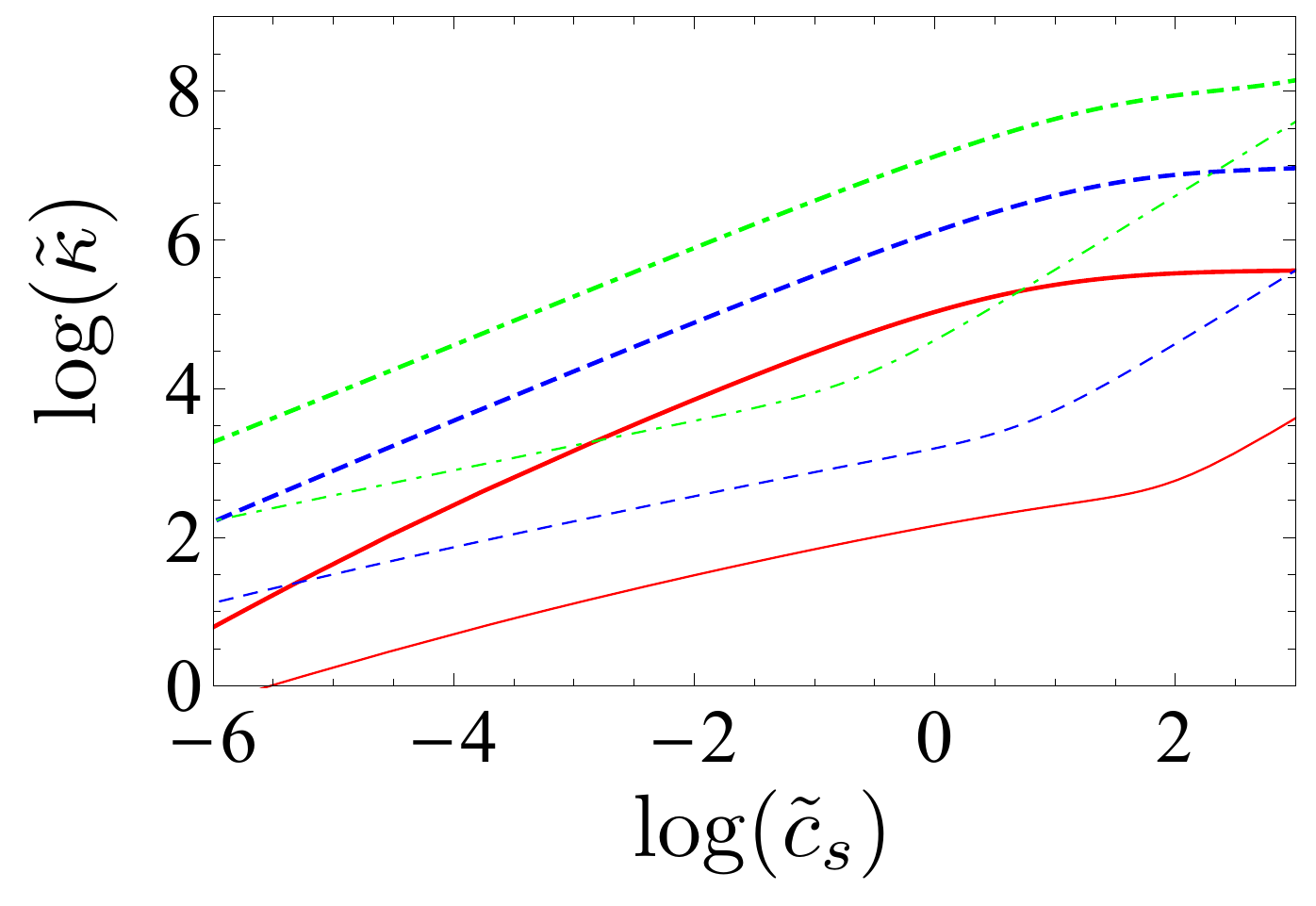}
\end{center}
\caption{Dimensionless conductivity $\tilde \kappa$ vs. $\tilde c_s$ for $p{\rm H}-pK=3$, $\sigma_0 =6.8$~C/m$^2$, $b=30$~nm (thick curves), and $R=1$ (solid red), 10 (dashed blue) and 100~nm (dotted-dashed green) (Log-Log scale). The thin curves corresponds to the no slip case ($b=0$).}
\label{plots_R}
\end{figure}

In Figure~\ref{plots_R} is studied the influence of the nanopore radius $R=1,10,100$~nm on the dimensionless conductance $\tilde G=\tilde \kappa$ for $b=1$ and 30~nm. It clearly shows that the smaller the pore, the more influent the slippage is. Moreover for large $b$, increasing $R$ only shifts the conductance to higher values without changing its shape. Keeping $\sigma$ and $c_s$ fixed, $\tilde c_s/\sigma^*\propto R$ which shows that it becomes increasingly difficult to enter the GCE regime ($\tilde c_s/\sigma^* < 1$) and therefore observe the plateau and scaling behaviors for large pores.

\section{Discussion}

Equations (\ref{KCRM}) and (\ref{CR}) are quite general and should therefore apply to any cylindrical nanopore whatever its chemical composition. In particular, \eq{KCRM} has been successfully used to fit conductivity measurements of NaCl in polymeric track-etched nanopores with radii varying from 0.5 to 5~nm~\cite{SciRep}. In these experiments, the conductance clearly shows a plateau at low concentrations, $c_s<10^{-2}$~mol/L, which is the signature of a constant surface charge density. Since these nanopores were coated with hexamethyldisilazane, their surface was hydrophobic, as confirmed by MD simulations~\cite{SciRep}. These nanopores turn out to be very weakly charged, with fitted values $0.05<\sigma^*<0.4$ and $b\simeq30$~nm. Hence the electro-osmotic contribution due to flow slippage at the nanopore surface plays a dominant role [last term on the rhs of \eq{KCRM}].

\begin{figure}[t]
\begin{center}
\includegraphics[width=6cm]{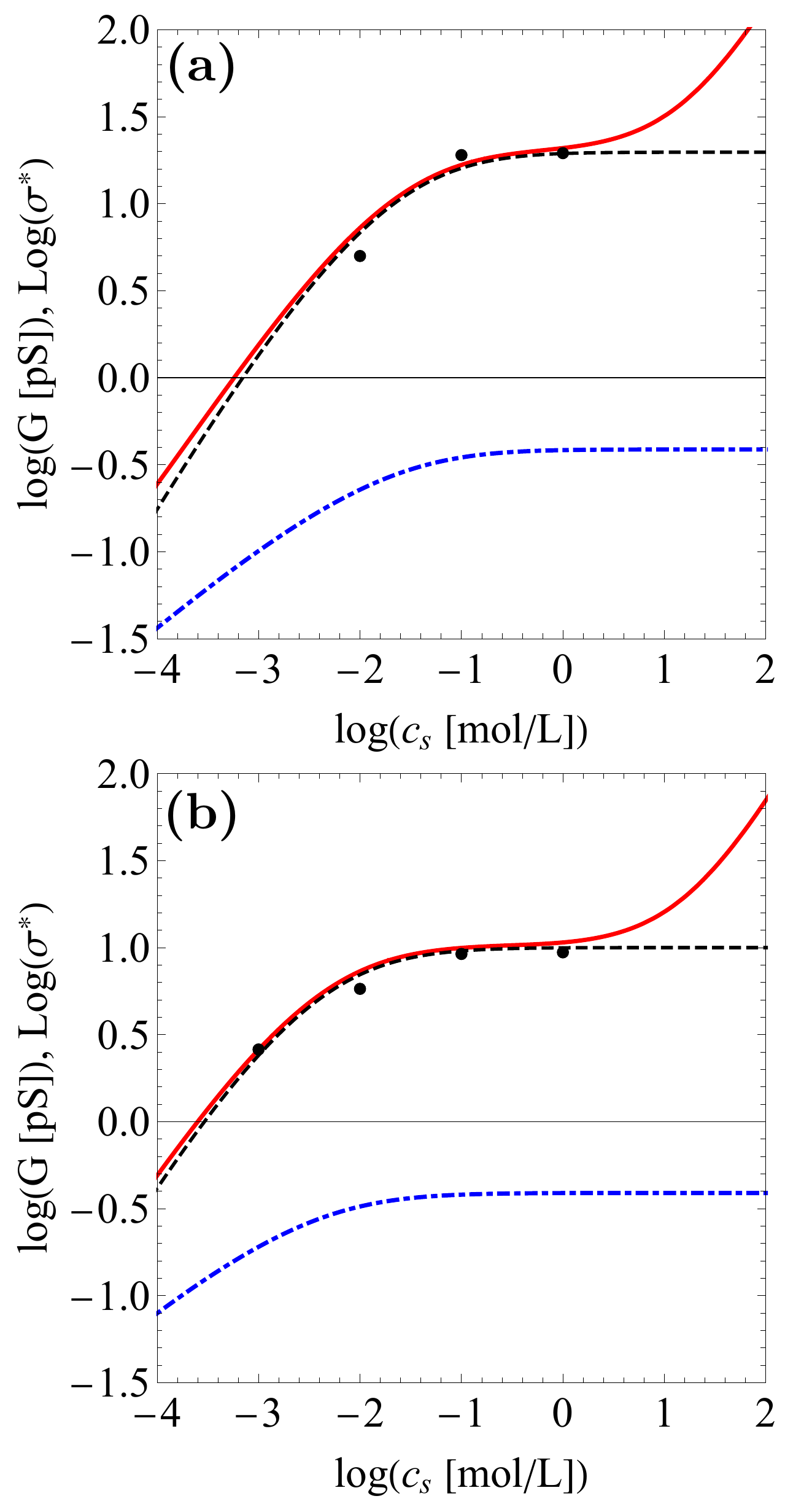}
\end{center}
\caption{Conductance (KCl) of devices with SWCNT versus the reservoir salt concentration $c_s$: (a)~device with a single nanotube ($R=0.7$~nm, $L=40~\mu$m), (b) device with 8 nanotubes ($R=0.7$~nm, $L=20~\mu$m). The red solid line is the fit using \eq{kappa}, and the black dashed line corresponds to the slip contribution, $\propto 4b\sigma^2$. The surface charge density $\sigma^*(c_s)$ is also shown (blue dotted-dashed line). Fitting parameters are $\sigma^*_0=0.39$, $\tilde b=71$, and $pK=4.92$ (a) and 4.17 (b).\label{Khadija}}
\end{figure}
Experiments on single-walled CNTs have been performed recently for radii between 0.6 and 1~nm and using KCl~\cite{Yazda2017}. For those that showed a linear $I-V$ curve, the conductance was measured as a function of $c_s$. The results are reproduced in Fig.~\ref{Khadija} together with the fits (in red) using \eqs{KCRM}{CR} (the mobility of K$^+$ is $\mu_+=5.2\times10^{11}$~s/kg). Is also shown the surface charge density versus $c_s$ (dotted-dashed blue curve). Figure~\ref{Khadija}(a) corresponds to a device with one unique nanotube.  For the other device [Figure~\ref{Khadija}(b)], 8 tubes are present but the number of conducting tubes is unknown. To get approximately the same conductance value as for the preceding device, we assumed that 5 tubes were conducting.  The fits are reasonably good, the fitting parameter values (for the 2 devices) being $b=50$~nm, $\sigma_0=0.041$~C/m$^2$ and $pK=4.92$ for (a) and 4.17 for (b) (for $p{\rm H}=7$ in the experiments). These slight differences can be due to variation of the $p$H from one sample to the other or slight differences in the nature of surface charges. The plots clearly show that the slip contribution given in in the last term of \eq{LSL}dominates the nanopore conductivity in the salt concentration range of interest (black dashed line). Moreover,  charge regulation is absolutely necessary to reproduce this sub-linear dependence, which interpolates between the asymptotic very low concentration law in $c_s^{1/2}$ [electrical migration (first term in~\eq{LSL}) not seen in the plots] and the plateau corresponding to surface charge saturation, by first passing through an intermediate scaling linear in $c_s$ (second term in \eq{LSL} dominated by slip).
The nanopore is weakly charged since $\sigma_0^*=0.39$, which puts this system in the homogeneous regime. This behavior is therefore qualitatively similar to that observed for the dashed-dotted red curve in Fig.~\ref{diag}.

\begin{figure}[t]
\begin{center}
\includegraphics[width=7cm]{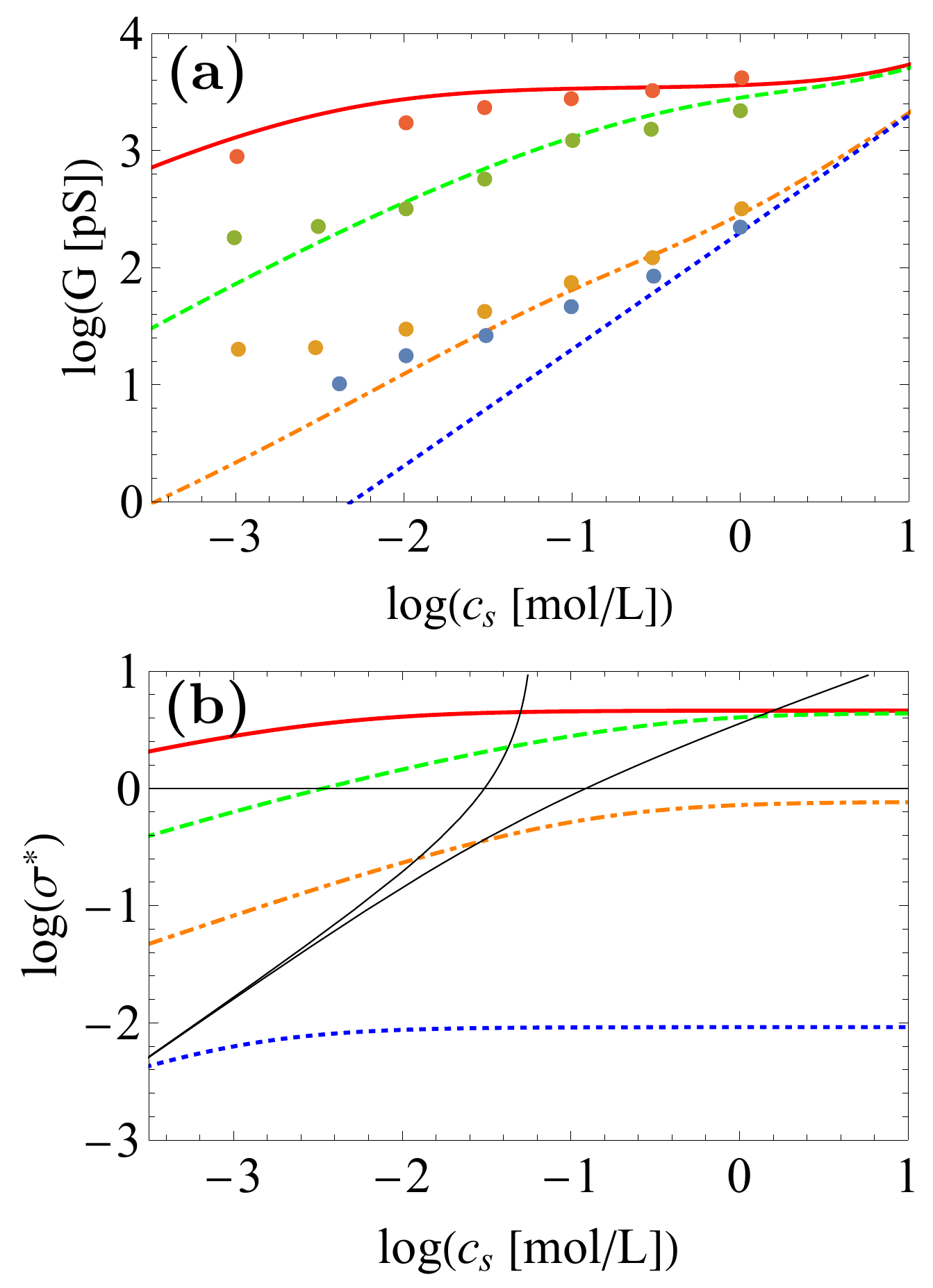}
\end{center}
\caption{(a) Conductance (KCl) measured in CNTs (data from Secchi \textit{et al.}~\cite{Secchi2016}) and fits using \eqs{KCRM}{CR} ($R=3.5$~nm, $L=3~\mu$m and $p{\rm H}=4,6,8,10$ from bottom to top). Fitting parameter values are: $pK=6.69$, $\sigma_0^*=4.6$ and $\tilde b=13$. (b)~ Associated (dimensionless) surface charge density $\sigma^*(\tilde c_s)$. The thin solid black lines correspond to $c_{\rm GCE}(\sigma^*)$ (upper) and $c_{\rm bulk}(\sigma^*)$ (lower). \label{LB}}
\end{figure}
Recently Secchi \textit{et al.}~\cite{Secchi2016} reported experimentally and proposed theoretically that for various nanopore radii ($R=3.5,10,14,35$~nm) and $p$H (from 4 to 10) the conductances of individual CNTs exhibit a power law $c_s^{1/3}$ behavior at low $c_s$. Note that, compared to the CNTs used in Ref.~\cite{Yazda2017}, these CNTs have  much larger radii, and are multi-walled.
To model their data, Secchi \textit{et al.} simplified the problem by adopting several assumptions (including the neglect of electro-osmosis) that enable them to uncover the intermediate scaling regime in $c_s^{1/3}$, which we have shown to be visible only for very highly charged nanopores in the absence of slip.

Biesheuvel and Bazant have fitted the data of Secchi \textit{et al.} by using the full space-charge/charge regulation model and solving numerically the PNP equations, but without slippage~\cite{Biesheuvel2016}. Although they succeeded in fitting the large radii ($R=10,14$ and 35~nm) conductivity data simultaneously for the four $p$H values studied using a physically reasonable surface charge density, they did not succeed in fitting the smallest pore radius ($R=3.5$~nm) (the model predictions deviate considerably from experiment for the two highest $p$H values).

Using our model without slip, we also obtained an acceptable fit for $R=3.5$~nm (not shown), but with $\sigma^*=500$, i.e. $\sigma_0=10.4$~C/m$^2$, which is unrealistic. In contrast, by taking into account slippage at the nanopore surface, we obtained a reasonable fit as shown in Fig.~\ref{LB}(a) except for the points at low $p$H and low $c_s$. The fitting parameters values are $pK=6.69$, $\sigma_0=0.095$~C/m$^2$ ($\sigma_0^*=4.6$) and $b=45.5$~nm which are all reasonable values. In particular the slip length is quite similar to the one used in Fig.~\ref{Khadija}. The surface charge density computed using this charge regulation model is shown in Fig.~\ref{LB}(b) versus  $c_s$. It shows that at $p{\rm H}=4$ (dotted blue curves) the surface is practically uncharged and in the experimental concentration range we expect bulk-like behavior. For $p{\rm H}=6$ (dotted-dashed orange curves) and 8 (dashed green curves) at low enough concentrations, we observe the weak charge GCE regime with what appears to be a 2/3 power law because of the combined contributions of electrical migration (1/3 power law) and electro-osmosis (1 power law), see \eq{LSL}. At $p{\rm H}=8$ and intermediate concentrations we also observe a 2/3 power law, but now because the system is in the strong charge GCE regime where slip dominates, see \eq{inter}. At $p{\rm H}=10$ (solid red curves) the surface charge density nearly saturates to $\sigma_{\max} \approx \sigma_0$ already at $c_s= 0.01$~mol/L. The experimental concentrations do not, however, reach low enough values to see clearly the low concentration scaling regime, just the cross-over from an incipient constant surface charge plateau to intermediate scaling in the high surface charge GCE regime, see \eq{inter}.

For the low $p{\rm H}=4$ value, the surface remains practically uncharged, which leads to the classical bulk behavior for the fitted conductance $G\propto c_s$, whatever the value of $c_s$. The data show a weaker slope (close to 1/3, as shown by Secchi \textit{et al.}), which can be obtained for a higher value of $p{\rm H}-pK\simeq 0$ [as shown in Fig.~\ref{th_plots}(b)], i.e. $pK\simeq4$. However this would imply a much higher, probably unphysical, surface charge density for the highest $p{\rm H}$ value, which leads to a very poor fit for the three other sets of data.

One reason why we obtain a too low conductance at low $c_s$ might be that some fixed charges remain on the nanopore surface at low $p$H, i.e. charged groups that cannot be neutralized by protons over the studied $p$H range (possessing for example a p$K\ll 4$) or surface-trapped charges due to doping. This is quite easy to implement in our model, by adding a residual negative fixed surface charge density of amplitude $\sigma_{\rm f}$ to the rhs of \eq{CRM2}. Equation~(\ref{CR}) is therefore modified according to
\be
\tilde c_s =\frac{ 2h\sigma^*(1+\sigma^*)(\sigma^*-\sigma_{\rm f}^*) (\sigma_0^*+\sigma_{\rm f}^*-\sigma^*) }{ [\sigma_0^*- (1+h) (\sigma^*-\sigma_{\rm f}^*)] [\sigma_0^*- (1-h) (\sigma^*-\sigma_{\rm f}^*)] }
\label{CR2}
\ee
and the parametric plot is done by enforcing that $\sigma_{\rm f}\leq\sigma\leq\sigma_{\rm f}+\sigma_{\max}(p{\rm H})$.
\begin{figure}[t]
\begin{center}
\includegraphics[width=7cm]{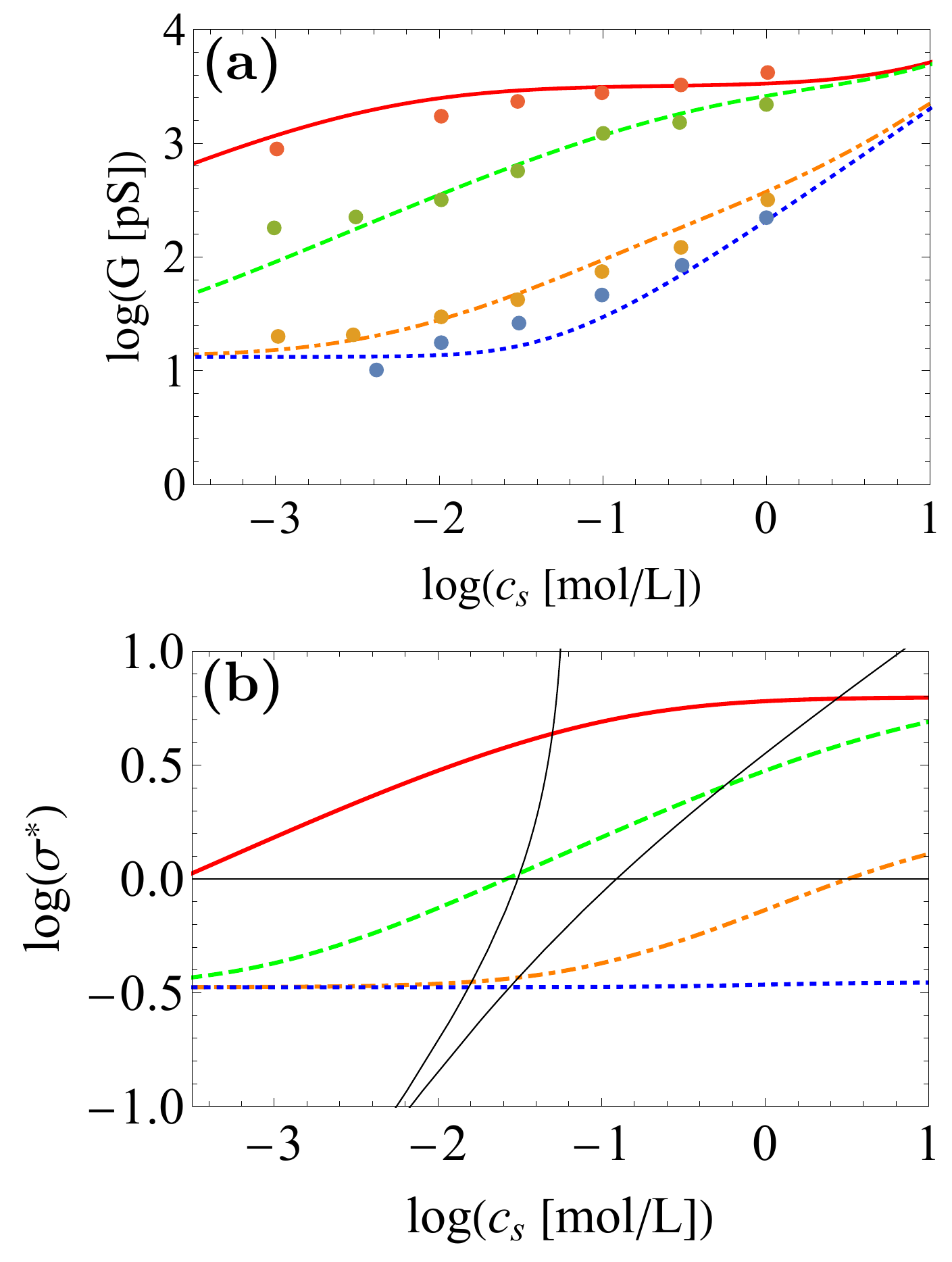}
\end{center}
\caption{Same plots as in Fig.~\ref{LB} but with an additional residual surface charge density $\sigma_{\rm f}$, i.e. fits are done using \eqs{KCRM}{CR2} Fitting parameter values are: $pK=6.53$, $\sigma_0^*=5.96$, $\tilde b=6.35$, and $\sigma_{\rm f}^*=0.33$.\label{LB2}}
\end{figure}
The corresponding plot is shown in Fig.~\ref{LB2}. The fit is much improved for low $p$H (dotted blue and dotted-dashed orange curves), showing a plateau at very low $c_s$. This plateau is due to the weak fixed surface charge density $\sigma_{\rm f}=6.9\times10^{-3}$~C/m$^2$, whereas $\sigma_0=0.124$~C/m$^2$. The fits at high $p$H are therefore not affected by this residual charge. The $pK=6.53$ is almost identical to the one found for $\sigma_{\rm f}=0$ whereas $b=22$~nm is slightly smaller. More data are needed at $p{\rm H}\leq 6$ and $c_s\simeq 10^{-3}$~mol/L to confirm the existence of such a residual surface charge.\\

Note that in this paper we assumed that the nanopore is uniformly charged. The eventual presence of charge defects leads to non-linear $I-V$ curves such as observed in ~\cite{Yazda2017}. Furthermore, we adopted the bulk values for the ionic mobilities and treated the ions at the mean-field level (thereby neglecting the effects of excluded volume, ionic correlations, and dielectric exclusion~\cite{PRL,JCP}). Finally we decided not to introduce a slip length that varies with the surface charge, but we have checked that the formula proposed by Huang \textit{et al.} in Ref.~\cite{Huang2008} and fitted from molecular dynamics simulations does not change the results. Secchi \textit{et al.} have shown recently~\cite{Secchi2016b} that the slip length in CNTs increases very quickly when the pore radius decreases from 50 to 15~nm, which will accentuate the increasing influence of slip for decreasing pore radius (see Fig.~\ref{plots_R}).
However a precise (experimental) law is still missing for smaller radii. Should this law be obtained in the near future, it would be easy to implement it in our theory.\\

In conclusion, a relatively simple analytical formula has been derived for the conductance in nanopores. It combines the classical bulk transport equation incorporating electrical migration with the electro-osmotic contributions. This last  contribution may become important in small nanopores with high surface charge and/or large slip length, which is the especially case  in CNTs. The charge regulation model is also solved analytically using an approximate formula  appropriate for cylindrical nanopores that interpolates between the homogeneous regime (described by the Donnan potential, valid for sufficiently weak surface charge density), the exact good co-ion exclusion limit (valid for sufficiently low salt concentration), and bulk behavior (valid for sufficiently high salt concentration). This formula allows us to extract the $pK$, the saturation surface charge density $\sigma_0$ and the slip length $b$ from experimental conductance measurements in track-etched nanotubes and CNTs. In particular a large variety of apparent exponents $\alpha$ governing the scaling regime of the conductance, $G\propto c_s^\alpha$, are observed at low and  intermediate salt concentrations (between $10^{-6}$ and 0.1~mol/L for nanopores), i.e. the experimental range of interest. For small diameter nanopores without flow slippage we find $\alpha\leq1/2$, in agreement with~\cite{Biesheuvel2016}), whereas for hydrophobic nanopores at high $p$H with strong slip effects, $1/2\leq\alpha\leq 1$ before a plateau ($\alpha=0$) at intermediate concentrations. At low $p$H the conductance crosses overs smoothly from the $\alpha = 1/2$ scaling at low concentration to a bulk behavior, $\alpha=1$, at high salt concentration.
For small radius nanopores, the $\alpha = 1/3$ scaling proposed in~\cite{Secchi2016} is uncovered as an intermediary cross-over regime that is only visible for unphysically high surface charge densities. It would be interesting if experiments could clearly detect the various accessible scaling regimes, especially the strong slip one and its qualitative signature, namely that at sufficiently high $p$H the conductivity approaches the asymptotic low concentration scaling ($\alpha = 1/2$) from above in the presence of slip and below without (see Fig.~\ref{th_plots}).
In order to fit the data for the 3.5~nm radius CNT of \cite{Secchi2016} using physically reasonable surface charge densities, we have shown that is necessary to include not only slip, but also a fixed surface charge (in addition to a charge regulation contribution).

We hope that the relatively simple analytic approach proposed here will help experimentalists not only to better characterize their nanopores, but also to better plan their experiments and ameliorate their nanopore design protocols. After more high quality conductivity measurements for CNTs are modeled using our approach, a clear assessment can be made to determine if in this case the Space Charge model needs to be extended beyond the present model. A key open question concerns the importance of the dielectric,  ion correlation, and excluded volume effects mentioned above, which provide corrections to the mean field Space Charge model, on ionic conductivity in nanopores, especially when coupled to charge regulation (for work in this direction see~\cite{JCP1,JCP}).
If there is experimental evidence for the need for further extensions, the important recent modeling work discussed in the introduction could also provide significant contributions~\cite{Maduar,Bonthuis12,Bonthuis13,Wang,Bieshpp16}).

\textbf{Note added:} While this article was under revision, an article citing our work (preprint~\cite{Manghi}) was submitted and published~\cite{Uematsu}. This recent modeling work, which included charge regulation, but not slip, reproduced our scaling analysis without slip and fully corroborated it by solving numerically the full Space Charge Model. When applied to the CNTs studied in~\cite{Secchi2016}, this approach led to good simultaneous fits to the conductivity data for the 35~nm radius nanopore as a function of salt concentration for the four $p$H values studied, but not for the 3.5~nm radius one at $p$H 6, despite using an unphysically high maximum surface charge density (a maximum ionizable surface site density of 19/nm${}^2$, equivalent to a maximum surface charge density of 3~C/m${}^2$, obtainable at high pH). This value, which is greater than the one obtained for extremely highly charged BNNTs~\cite{Siria2013}, does not appear to be compatible with what is known about CNTs~\cite{Skwarek}.  Fig.~2d of~\cite{Biesheuvel2016} shows that if physically reasonable values of maximum surface charge density are used without slip the situation is even worse for the 3.5~nm radius nanopore (the data for the two highest $p$H values are poorly fitted).

\acknowledgments
This work was supported in part by the French Research Program ANR-blanc, Project TRANSION (ANR-2012-BS08-0023).

\appendix*
\section{General result for the slip contribution to conductivity}

In this appendix, we show that the slip contribution to conductivity is, within the scope of the PNP model, an additional contribution equal to
\be
\kappa_{\rm slip}=\frac{2\sigma^2b}{\eta R}
\label{slip2}
\ee
as given in \eq{slip} (see SI for~\cite{SciRep}).
The Stokes equation along the axial direction $z$ is
\be
\frac{\eta}{r}{\partial_r}\left(r\partial_r v_z\right)-\rho_c\partial_zV-\partial_zp=0
\ee
where $V(z)=-z E$ arises from the applied voltage difference, $\Delta V=-L E$ (where $L$ is the length of the nanopore, $p$ is the pressure, and the charge density $\rho_c$ is related to the electrostatic potential $\phi (r)$ through the Poisson equation
\be
\rho_c=-\frac{\epsilon_0\epsilon}{r} \partial_r(r\partial_r \phi).
\label{Poisson}
\ee
Inserting \eq{Poisson} in \eq{slip2} and using the slip boundary condition
\be
v_z(r)+b\partial_rv_z(r)|_{r=R}=0
\ee
and the Gauss law $\partial_r\phi|_{r=R}=\sigma/(\epsilon_0\epsilon)$, yields the modified Helmholtz-Smoluchowski equation
\be
v_z(r)=-\frac{\epsilon_0\epsilon}{\eta}[\phi(R)-\phi(r)]\partial_z V-\frac{\partial_z p}{4\eta}(R^2-r^2)+v_{\mathrm{slip},z}
\ee
where the slip velocity is
\be
v_{\mathrm{slip},z}=\frac{b}{\eta}\left(\sigma\partial_z V-\frac{R}2\partial_z p\right)
\ee
Since the slip velocity is a constant, the advective contribution to the pore averaged electric current is directly
\be
J_{\mathrm{slip},z}=v_{\mathrm{slip},z}\langle\rho_c(r)\rangle=-2\frac{b\sigma}{\eta R}\left(\sigma\partial_z V-\frac{R}2\partial_z p\right)
\ee
where the electroneutrality, $\langle\rho_c(r)\rangle=-2\sigma/R$, in the pore has been used. Hence the slip contribution to conductivity, defined by $\kappa_{\rm slip}=-J_{\mathrm{slip},z}/\partial_z V$ for $\partial_zp=0$, is given by \eq{slip2}.

\end{document}